\documentclass[twocolumn,noshowpacs,nofootinbib,10pt]{revtex4-1}

\usepackage{float,xcolor,upgreek,yfonts,tikz}
\usepackage{color}
\usepackage{diagbox}
\usepackage{graphicx}
\usepackage{graphicx,float}\usepackage[all]{xy}
\usepackage{amsmath,upgreek,bigints}
 \newcommand{\bltx}{\textcolor{black}}
 \usepackage{caption}
 \usepackage{booktabs,makecell}
\usepackage{subcaption}
  \newcommand{\blt}{\textcolor{black}}
\usepackage{amssymb}

\usepackage{alphabeta}
\usepackage{textgreek}

\newcommand{\beq}{\begin{eqnarray}}
\newcommand{\eeq}{\end{eqnarray}}
\newcommand{\be}{\begin{equation}}
\newcommand{\ee}{\end{equation}}

\newcommand{\bea}{\begin{eqnarray}}
\newcommand{\eea}{\end{eqnarray}}

\newcommand{\ba}{\begin{eqnarray}}
\newcommand{\ea}{\end{eqnarray}}
 \def\de{\partial}

\usepackage[colorlinks,hyperindex,unicode]{hyperref}
\definecolor{green1}{RGB}{0,128,0} 
\hypersetup{hidelinks,backref=true,pagebackref=true,hyperindex=true,colorlinks=true,breaklinks=true,urlcolor= blue}
\hypersetup{%
  colorlinks = true,
  linkcolor  = blue,
  citecolor = cyan,
}
\usepackage{bookmark,textgreek}
\usepackage{hyperref,color,xcolor}
\hypersetup{hidelinks,hyperindex=true,colorlinks=true,breaklinks=true,urlcolor= blue}
\hypersetup{%
  colorlinks = true,
  linkcolor  = blue
}
\newcommand\orcidroldao{{\href{https://orcid.org/0000-0003-3978-532X}{\orcidicon}}}
\newcommand{\orcidicon}{%
	\begin{tikzpicture}
	\draw[lime, fill=lime] (0,0)
		circle [radius=0.16]
		node[white] {{\fontfamily{qag}\selectfont \tiny ID}};
	\draw[white, fill=white] (-0.0625,0.095)
		circle [radius=0.007];
	\end{tikzpicture}	\hspace{-2mm}
}

\newcommand{\bse}{\begin{subequations}}
\newcommand{\ese}{\end{subequations}}

\input amssym.def
\input amssym.tex

\begin{document}
\title{Holographic entanglement entropy, deformed black branes and deconfinement in AdS/QCD}

\author{Roldao da Rocha\orcidroldao\!\!}
\affiliation{Center of Mathematics, Federal University of ABC, 09210-580, Santo Andr\'e, Brazil.}
\email{roldao.rocha@ufabc.edu.br}

\begin{abstract} 
A family of deformed black branes is employed to examine the confinement/deconfinement phase transition in AdS/QCD. The holographic entanglement entropy (HEE) plays the role of the order parameter driving the confinement/deconfinement phase transition. The binding energy of quark-antiquark bound states and the critical length that makes a null variation of the HEE, together with the critical temperature separating the deconfined quark-gluon plasma to the confined hadronic phase, are discussed in the deformed black brane background, matching current experimental hadronic data. 

\end{abstract}

\pacs{04.50.Gh, 12.38.Aw, 03.67.-a}

\keywords{AdS/QCD; holographic entanglement entropy; black branes; quarks; deconfinement phase transition}

\maketitle

\section{Introduction}
\label{111}

The deconfinement phase transition, above which quark bound states occupy a deconfined phase and one may observe individual quarks, has been not completely grounded onto the QCD framework yet. Despite current experimental data, a thorough theory that unfolds QCD at arbitrary energy ranges is far beyond existing analytical methods, although lattice QCD can reasonably scrutinize non-perturbative issues. At the QCD ultraviolet regime, asymptotic freedom permits using perturbative techniques that may yield QCD to match low-energy experimental data \cite{Aharony:1999ti}. Nevertheless, QCD underlies the quarks and gluons paradigm, whereas experimental data at low energies reckons bound states constituting hadrons. Besides, QCD is endowed with an inherent nonlinearity due to self-interactions among gluonic degrees of freedom. \blt{At the strong coupling regime, perturbative techniques do not encompass all possible scenarios \cite{Witten:1979kh}.} \blt{Although the QCD setup based on the perturbative expansion
 in $\alpha_s(Q^2)$ cannot be applied
for the description of physical processes at small $Q^2$, where the running strong
coupling constant has order $\mathcal{O}(1)$, one can still apply the large $N_c$ QCD as a perturbative technique.}
In particle collision processes at high energy, color electric flux tubes link quarks ($q$) to antiquarks ($\bar{q}$), which can also shred apart into either baryons and mesons or gluons and quarks altogether \cite{Greensite:2003xf}. 
Contemporary progresses, mainly in AdS/QCD, have shed new light on hadronic constituents \cite{Brodsky:2007hb,Karapetyan:2018yhm,Karapetyan:2018oye,BallonBayona:2007vp,Ferreira:2020iry,Ferreira:2019nkz,Ali-Akbari:2017vtb}. 
\blt{Since QCD is strongly coupled at the low-temperature regime, perturbative techniques at low-temperature regime approaching the deconfinement phase transitions have been developed. From the gravity dual side, deconfinement phase transitions correspond to the transition between an AdS${}_5$-Schwarzschild black brane and a pure AdS${}_5$ spacetime.} \blt{The influential Refs. \cite{Gutsche:2019blp,Gutsche:2019pls,Gutsche:2019jzh} verified and demonstrated the correspondence with the temperature-dependence in Chiral Perturbation Theory (ChPT). The effective expansion parameter is given by $R = \frac{T^2}{12F^2}$, where $F$ is the leptonic decay constant, which is roughly 100 MeV. It is clear that for the typical value of the critical temperature $T_c = 155$ MeV, the expansion parameter reads $R\approxeq 0.2$, which is reasonably small to perform the perturbative expansion up to and including the confinement/deconfinement phase. Other important developments were presented in Refs. \cite{Lyubovitskij:2020gjz,Lyubovitskij:2020xqj}. 
Also, non-perturbative methods can investigate the deconfinement phase transition, addressing its temperature and its order parameter as well \cite{Boschi-Filho:2002glt}}. Besides, 
an appropriate setup also includes determining the screening range of the $q\bar{q}$ potential, as well as their binding energy.

Gauge/gravity duality is an effective non-perturbative procedure that implements the correspondence between a $d$-dimensional strongly coupled quantum Yang--Mills theory and weakly coupled gravitational configurations in $d+1$ dimensions \cite{Witten:1998qj,Maldacena:1997re}. 
The dual gravity scenario can successfully emulate QCD phenomenology, in the AdS/QCD approach. 
In this scenario, entanglement entropy (EE) can be also computed in several contexts to study observables in QCD \cite{Nishioka:2009un}. 
The EE quantifies the correlations
between subsystems in some larger network described by quantum mechanics. For two subsystems, split apart by a surface, the EE is proportional to the surface area and may also be controlled by an ultraviolet cutoff regulating correlations at short distances \cite{Solodukhin:2011gn}. 
The holographic EE (HEE) is a very relevant instrument to scrutinize several aspects of AdS/CFT, particularly AdS/QCD and its phenomenology \cite{Ryu:2006bv}. Entangled quantum fields in QCD can be studied from the weakly coupled dual gravitational systems  
\cite{Hubeny:2007xt,Emparan:2006ni}. 
Ref. \cite{Klebanov:2007ws} used the HEE to probe a deconfinement phase transition at the zero temperature regime. Phase transitions involving the finite temperature case were implemented in Refs. \cite{Knaute:2017lll,Dudal:2018ztm}. Extended anisotropic black hole solutions on fluid branes were investigated from the point of view of the HEE \cite{daRocha:2020gee,DaRocha:2019fjr}, being employed to analyze hot hadronic media \cite{Bittencourt:2015ejb,Bittencourt:2018ywl}. 

In this work, the HEE underlying AdS/QCD will be explored as an order parameter controlling the deconfinement phase transition, in a deformed black brane scenario. Several aspects of the deconfinement phase transition will be addressed and discussed. The $q\bar{q}$ binding energy, the deconfinement phase transition critical temperature at which a transition from hadronic matter to quark-gluon plasma occurs, and the critical length that comprises the separation between a quark and an antiquark in a $q\bar{q}$ system will be analyzed in the light of the AdS/QCD hard and soft wall models and current experimental data as well. The parameter that defines a family of deformed black branes will be shown to lie in a range that is stricter than the bound obtained by the shear viscosity-to-entropy density ratio. This range will be derived when considering the HEE and the critical temperature at which the deconfinement phase transition occurs, matching current experimental data in QCD. 
This paper is organized as follows: Sec. \ref{c21} is devoted to presenting a 1-parameter family of deformed black branes, and associated thermodynamical quantities. The shear viscosity-to-entropy density ratio determines the range of the deformed black brane parameter and the holographic Weyl anomaly is discussed. In Sec. \ref{c23}, the confinement/deconfinement phase transition in the QCD dual gauge theory at the AdS boundary is explored. The HEE is presented and addressed as the order parameter controlling the confinement/deconfinement phase transition. The binding energy of the $q\bar{q}$ bound state and the critical length that makes a null variation of the HEE, together with the critical temperature separating the deconfined quark-gluon plasma to the confined hadronic phase, are analyzed in the deformed black brane setup, matching current experimental hadronic data. Reciprocally, QCD phenomenology drives a more strict range for the deformed black brane parameter. Sec. \ref{c34} is dedicated for further discussion, concluding remarks, and perspectives of the relevant results here obtained.

\section{Deformed Black Branes}
\label{c21} 
\begin{figure}[H]
\begin{center}
\includegraphics[width=76 mm]{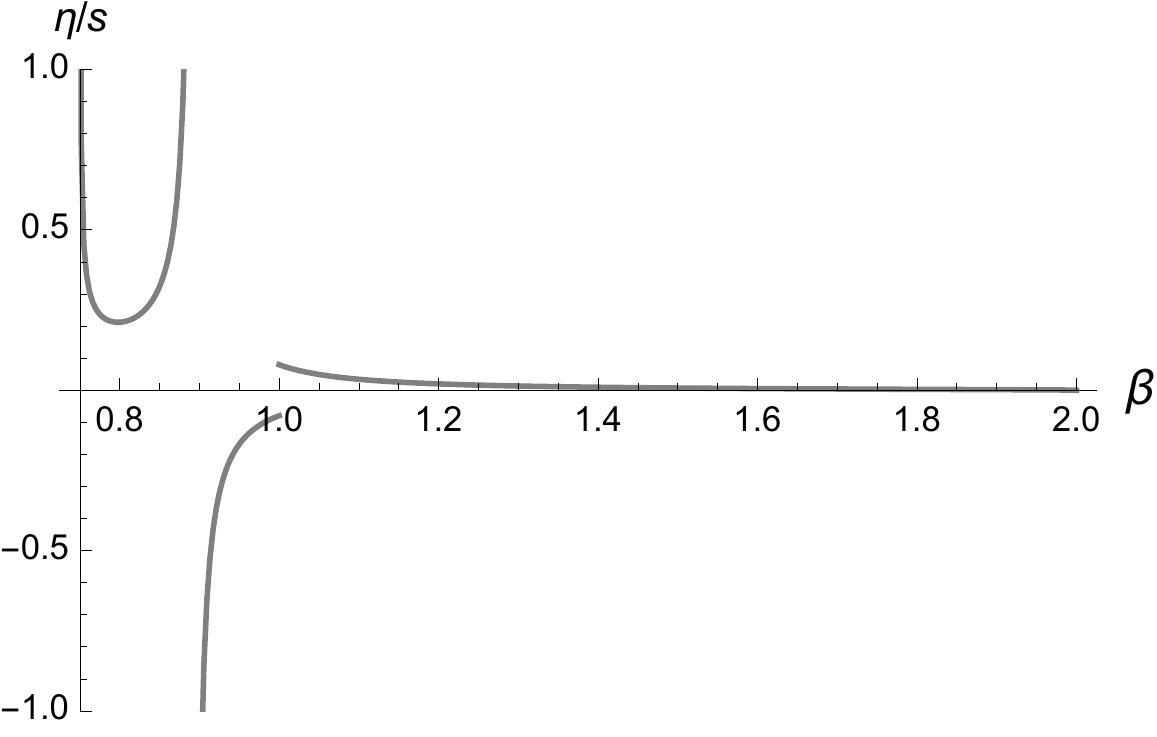}
\caption{\small{$\eta/s$ as a function of $\upbeta$. \label{etas}}}
\end{center}
\end{figure}

A family of deformed black branes was derived and discussed in Ref. \cite{Ferreira-Martins:2019svk}, using AdS/CFT and the ADM and Hamiltonian constraints. Deformed black branes generalize the standard AdS$_5$--Schwarzschild black brane and play a prominent role in gauge/gravity duality.
The deformed black brane has metric given by
\begin{eqnarray}\label{metric1}
ds^2 = \frac{R^2}{z^2} \left(-N(z) \mathrm{d}t^2 + \updelta_{ij} \mathrm{d}x^i \mathrm{d}x^j + \frac{1}{C(z)} \mathrm{d}z^2\right), 
\end{eqnarray}
where
\begin{eqnarray}
N(z) &=& 1 - \frac{z^4}{z_0^4} + \left (\upbeta - 1 \right ) \frac{z^6}{z_0^6},\label{eq:Nu}\\
C(z) &=& \left (1 - \frac{z^4}{z^4_0} \right ) \left ( \frac{2 - \frac{3z^4}{z_0^4}}{2- \left (4\upbeta-1\right ) \frac{z^4}{z_0^4}}\right).
\label{eq:Au}
\end{eqnarray}
with event horizon at $z_0$. 
%Given the deformed black brane geometry, bulk fields do correspond to Yang--Mills operators in the boundary, and the gravitational description can be employed to compute $n$-points correlation functions of the Yang--Mills operators in the QFT living in the boundary.
The standard AdS$_5$--Schwarzschild metric 
is recovered whenever $\upbeta\to1$. For fitting the deformed black brane to the slope of standard Regge trajectories in QCD, a conformal factor $e^{cz^2/2}$ is usually accounted in \eqref{metric1}, with $c \sim 0.9 \,{\rm GeV}^2$. 
When Eqs. (\ref{metric1}, \ref{eq:Nu}, \ref{eq:Au}) are regarded, the Gubser--Klebanov--Polyakov--Witten relation \cite{Witten:1998qj,Gubser:1998bc} yields the partition function associated with the dual theory at the boundary \cite{Ferreira-Martins:2019svk}, \begin{equation} \label{partfnc}
\!\!\!\!\bltx{Z\sim\frac{R^{4}}{16\pi G_5}\left({3\upbeta^{2}-15\upbeta+11}\right).}
\end{equation}
 The Hawking temperature at the deformed black brane horizon \cite{Ferreira-Martins:2019svk},
	\begin{eqnarray} \label{ttt}
	T=\frac{R}{\pi z_0}\sqrt{\frac{\upbeta-2}{3-4\upbeta}},
	\end{eqnarray}
 diverges at $\upbeta\to 0.75$, and attains imaginary values in the range $\upbeta\in(-\infty, 0.75)\cup (2,+\infty)$, yielding the allowed open range $\upbeta\in(0.75,2)$ for the deformation parameter. 
The free energy, the entropy density, the pressure, and the energy density can be immediately derived from the partition function \ref{ttt}, and are respectively given by \cite{Ferreira-Martins:2019svk} 
\begin{eqnarray} \label{entropyofT}
\label{freenergyofT}
	\!\!\!\!\!\!\!F&=&\frac{\pi^{3}V}{16G_5}\left({3\upbeta^{2}-15\upbeta+11}\right)\left(\frac{4\upbeta-3}{\upbeta-2}\right)^{2}T^{4},\\\!\!\!\!\!\!\!\!s&=&-\frac{R^{3}}{4G_5}\left({3\upbeta^{2}-15\upbeta+11}\right)\sqrt{\frac{4\upbeta-3}{\upbeta-2}},\\
			\!\!\!\!\!\!\!\!P&\!=\!&\!-\frac{\pi^{3}}{16G_5}\left({3\upbeta^{2}-15\upbeta+11}\right)\left(\frac{4\upbeta-3}{\upbeta-2}\right)^{2}\!T^{4}\ ,\label{pofT}\\
			\!\!\!\!\!\!\upepsilon &=& \!\frac{5\pi^{3}}{16G_5}\!\left({3\upbeta^{2}-15\upbeta+11}\right)\left(\frac{4\upbeta-3}{\upbeta-2}\right)^{2}\!T^{4}\label{espsofT}.
	\end{eqnarray}
Regarding a perfect fluid, Eqs. (\ref{pofT}, \ref{espsofT}) evaluated at the
boundary implies the trace of the energy-momentum tensor  
to read 
\beq
 g_{\mu\nu}T^{\mu\nu} & =&-\frac{\pi^{3}}{2G_5}\left({3\upbeta^{2}-15\upbeta+11}\right)\!\left(\frac{4\upbeta-3}{\upbeta-2}\right)^{\!\!2}T^{4}.\label{trEMT}
\eeq
The shear viscosity-to-entropy density ratio can be also derived,  
\begin{equation} \label{eq:etaSfinal}
\frac{\eta}{s}=\frac{1}{4\pi}\left(\frac{1}{3\upbeta^{2}-15\upbeta+11}\right)\sqrt{\frac{4\upbeta-3}{\upbeta-2}}.
\end{equation} 
Fig. \ref{etas} illustrates the profile of $\eta/s$ as a function of $\upbeta$.
\noindent Since the range $0.9\leq\upbeta < 1$ is clearly forbidden, as it implies that $\frac{\eta}{s}<0$, the deformation parameter can therefore lie in the range 
\begin{equation}\label{ra1}
\upbeta\in(0.75, 0.9)\cup(1,2].
\end{equation}
 The saturation $\frac{\eta}{s} = \frac{1}{4\pi}$ implies $\upbeta = 1$, recovering the Kovtun--Son--Starinets (KSS) seminal result for the AdS$_5$--Schwarzschild black brane. Ref. \cite{Ferreira-Martins:2019svk} showed that the existence of a real Killing event horizon implies a narrower range bound, $\upbeta\in(0.75, 0.9)\cup(1,1.234]$ \cite{Meert:2020sqv}. Besides, the holographic Weyl anomaly can be emulated for the deformed black brane, for $a$ and $c$ being central charges of the conformal gauge field theory, as  
 \beq\label{wa}
\!\!\!\!\!\!\!\! 16\pi^2\langle T^\mu_{\;\,\mu}\rangle_{\scalebox{0.54}{$\textsc{CFT}$}}&=& {c}{}\left(R^{\mu\nu\alpha\beta}R_{\mu\nu\alpha\beta}-2R^{\mu\nu}R_{\mu\nu}+\frac{R^2}{3}\right)\nonumber\\&&-{a}\left(R^{\mu\nu\alpha\beta}R_{\mu\nu\alpha\beta}\!-\!4R^{\mu\nu}R_{\mu\nu}\!+\!R^2\right),
 \eeq involving the Riemann tensor and its contractions.
 Running the calculations for the deformed black brane background (\ref{metric1}, \ref{eq:Nu}, \ref{eq:Au}), one obtains the expansion 
 \beq
\!\!\!\!\!\!\!\!\!\!\langle T^\mu_{\;\,\mu}\rangle_{\scalebox{0.54}{$\textsc{CFT}$}} \!=\! N^2\left[\frac{400}{3}\!+\!\frac{320}{3}(\upbeta-1)\frac{z^4}{z_0^4}+ 
80(\upbeta-1)\frac{z^6}{z_0^6}\right]
 \eeq
up to $\mathcal{O}(z^8)$, near the boundary, where $N^2=\frac{\pi L^3}{2G_5}$. Therefore, considering either $\upbeta\to 1$ or $z\to0$, the holographic Weyl anomaly associated with the AdS$_5$--Schwarzschild black brane can be recovered \cite{Ferreira-Martins:2019svk}.

\section{Holographic Entanglement Entropy and deconfinement phase transition} 
\label{c23}
The potential energy of a $q\bar{q}$ system can be computed when one takes into account Wilson loops in Yang--Mills theory, that can be derived from the gauge connection holonomy around a loop \cite{Boschi-Filho:2005nmp}. One can consider a loop ${\scalebox{0.9}{$\textsc{C}$}}$ of rectangular form, whose sides are the time coordinate, $t$, and the distance among confined quarks, $r$, with $r\ll t$. The expectation value of the Wilson loop, $\langle W({\scalebox{0.9}{$\textsc{C}$}}) \rangle =e^{-i(V(r)+2m)t},$ encodes the potential energy $V(r)$ associated with the $q\bar{q}$ pair, where $m$ denotes the quark mass. 
The holographic dual of the Wilson loop is the action $S$ of a string, whose endpoints are separated by a distance $r$, and can be written as \cite{Casalderrey-Solana:2011dxg}
\be %
\label{action}
 S=-\frac{1}{2 \pi \alpha'} \iint \sqrt{- \det(g_{rs})} \,d\uptau d\upsigma, 
\ee %
where $g_{rs}={\rm g}_{MN}\frac{\partial x^M}{\partial \zeta^r} \frac{\partial x^N}{\partial \zeta^s}$ is the induced metric on the string worldsheet, for $\zeta^1 = \uptau$ and $\zeta^2 = \upsigma$; $x^M(\zeta^r)$ stands for worldsheet coordinates, whereas ${\rm g}_{MN}$ denotes the background metric.
Therefore one can identify $
\langle W({\scalebox{0.9}{$\textsc{C}$}})\rangle =e^{i S({\scalebox{0.7}{$\textsc{C}$}})}.$ 

One can consider weakly coupled gravity in AdS${}_5$ as the dual theory to QCD. Coordinates in AdS${}_5$ are usually denoted by $(t,x_1,x_2,x_3,z)$, where $z$ is the energy scale in QCD. 
The action $S({\scalebox{0.9}{$\textsc{C}$}})$ can be computed, for $t=\uptau$ and $x_1=\upsigma$. The distance $r$ was calculated in Ref. \cite{Andreev:2006ct} as
\be\begin{split}
r=2\bigintsss_{\,0}^{z_\star}\frac{z^2e^{\frac{c}{2}\left(z_\star^2-z^2\right)}}{\sqrt{z_\star^4-z^4e^{c\left(z_\star^2-z^2\right)}}}\,dz,
\end{split}\ee %
where $\displaystyle{z_\star=\lim_{x_1\to0}z}$ plays the role of the string turning point. The range $cz_\star^2\in(0,2)$ corresponds to $r\in(0,+\infty)$. The upper limit $z_\star=\sqrt{2/c}$ was obtained in Ref. \cite{Andreev:2006ct}, where $c\approx 0.9$ GeV${}^2$.
 The potential energy in the QCD-like gauge theory, up to a multiplicative constant $\frac{b}{\pi z_\star^2}$, reads 
\be\label{potential} 
\!\!\!\!\!\!V(r)=\!\! \bigintsss_{\,0}^{z_\star}\!\! \left[\frac{1}{z^2}\left(\frac{e^{\frac{cz^2}{2}}}{\sqrt{z_\star^4\!-\!z^4e^{c\left(z_\star^2-z^2\right)}}}\!-\!1\right)\!-\!1\right]\,dz,
\ee 
where $b\approxeq 0.941$ matches experimental data \cite{Andreev:2006ct,Ali-Akbari:2017vtb}, 
yielding the expected linear profile at large distances and the $1/r$ regime for short distances.
In fact, the large distance limit is given by
\beq
\lim_{r\to0}V(r)=b\left(-\frac{\upkappa_0}{r}+k_0r+\mathcal{O}(r^3)\right),
\eeq
for $\upkappa_0\approxeq 0.23$ and $k_0\approxeq 0.16$ GeV${}^2$,
whereas the short distance regime is governed by 
\beq
\lim_{r\to\infty}V(r)=b k r,
\eeq
for $k\approxeq 0.19$ GeV${}^2$. 

In the deformed black brane scenario, the HEE setup can be implemented. Any quantum field theoretical system, at zero temperature, can be described by a pure lowest-energy state $| \psi \rangle$ and its associated density matrix, $\rho=| \psi \rangle\otimes\langle \psi |$. The quantum system can be split into two complementary subsystems $A$ and $B$, by a bipartition of the original Hilbert space {\scalebox{0.9}{$\mathcal{H}$}} = {\scalebox{0.9}{$\mathcal{H}_A\otimes \mathcal{H}_B$}}, when one looks at a spacelike subregion where an observer in $A$ accesses no degrees of freedom in $B$, implying that the (reduced) density matrix associated with $A$ is obtained by calculating the partial trace (tr) of the density matrix $\rho$, as $\rho_A={\rm tr}_B\rho$. In fact, given the state $|\psi\in \rangle \in {\scalebox{0.9}{$\mathcal{H}_A\otimes \mathcal{H}_B$}}$, the state of $A$ is the partial trace of over the basis of subsystem $B$, given by
\beq
\!\!\!\!\!\!\!\!\!\!\!\!\!\rho_A \!=\!\! \sum _{j}^{{\scalebox{0.65}{$\dim\mathcal{H}_B$}}}\!\!\left(I_{A}\otimes \langle j|_{B}\right)\left(|\psi \rangle\otimes \langle \psi |\right)\left(I_{A}\otimes |j\rangle _{B}\right)\!=\!{\hbox{tr}}_{B}\rho.\eeq
The EE of $A$ is given by the von Neumann entropy of $\rho_A$, namely, $S_A=-{\rm tr}_B(\rho_A\log\rho_A$) and quantifies how much information is lost when an observer is restricted to $A$, being isolated from $B$.

It codifies entanglement in the quantum information framework of the quantum field theory under scrutiny. From the gravitational dual side, the HEE, hereon denoted by $S_A$, can be computed by the expression \cite{Ryu:2006bv,Solodukhin:2011gn,Hubeny:2007xt,Emparan:2006ni,Nishioka:2009un} 
\be \label{eee}%  
 S_A=\frac{{\scalebox{0.84}{$\textsc{Area}$}}(\upgamma_A)}{4G_5},
\ee %
for $\upgamma_A$ denoting a codimension-2 minimal manifold, with boundary $\partial_{\upgamma_A} = \partial A$, in (asymptotically) AdS${}_5$, and $G_5$ is the bulk Newton's coupling constant. Also, $\upgamma_A$ must be homologous to the region $A$ and defined on the very same time slice as the region $A$. 
%The HEE $S_A$ represents the von Neumann entropy of the reduced density matrix when one spreads degrees of freedom inside a three-dimensional submanifold $B$ in a given 4-dimensional quantum field theory.
%$S_A$ encodes the correlation between $A$ and $B$, quantifying the entropy measured in $A$, by an observer to whom 
%$B$ is completely unaccessible. 
Since the HEE is divergent when the 
continuum limit is taken into account, an ultraviolet cutoff $a$ circumvents this divergence, whose coefficient is proportional to the area of
the boundary $\de A$. In this case one can write 
\be S_A\sim \frac{\mbox{{\scalebox{0.84}{$\textsc{Area}$}}}(\de
A)}{a^{2}}. 
 \label{divarea}\ee
It is worth mentioning that the cutoff is a necessary tool when addressing the 
Poincar\'e metric of AdS$_5$, with radius $R$,
\be ds^2=\frac{R^2}{z^2}\left(dz^2-dt^2+
dx_idx^i\right). 
\label{Poincare} \ee
At $z\to 0$ the metric (\ref{Poincare}) diverges, which can be circumvented by considering $z \geq a$ and making the boundary at $z=a$. 
In this setup, the HEE in the 4-dimensional conformal field theory can be computed by Eq. (\ref{eee}). Choosing $\upgamma_A$ to compute Eq. (\ref{eee}) is equivalent to determining the most solid 
entanglement entropy bound.
 The relationship between the HEE and the black hole entropy can be still established \cite{Nishioka:2009un}. 
When finite temperature sets in, dual strongly coupled plasmas can be described.

\begin{figure}[H]
\centering
 \hspace*{-0.9cm} \includegraphics[scale=0.352]{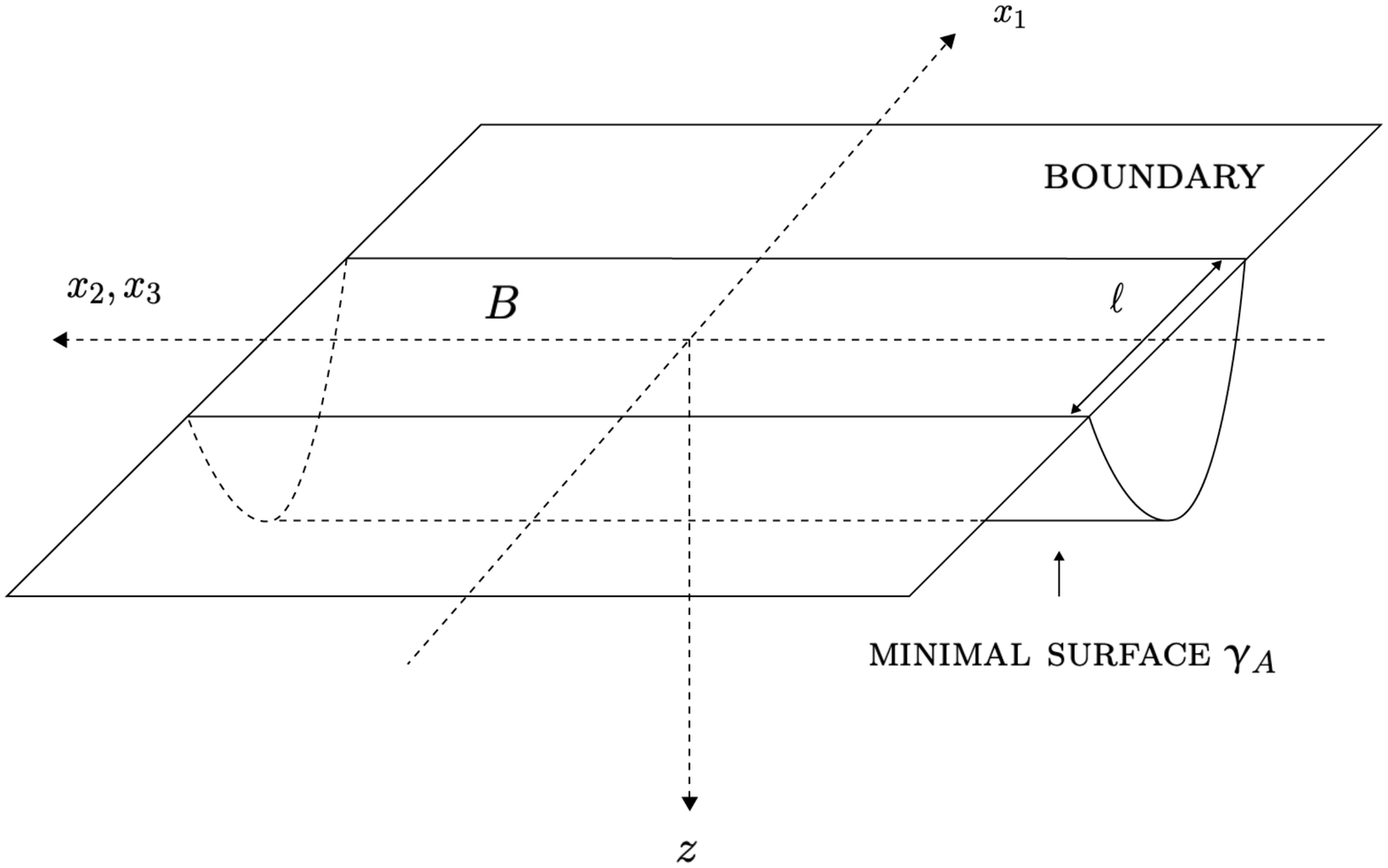}\vspace*{-0.6cm}
 \caption{\small{HEE of the region $A$: a hypercylindrical-type minimal surface, $\upgamma_A$, on the entangling region $A$, with surface area determining the HEE of the region $A$.}
}
 \label{fig:fig}
\end{figure}

The range for the parameter $\upbeta$ can be better refined, employing the HEE. The deformed black brane (\ref{metric1}, \ref{eq:Nu}, \ref{eq:Au}) has boundary that can be split into subsystems $A$ and $B$, where $B$ is defined by $-\frac{\ell}{2}<x_1<\frac{\ell}{2}$ and $x_2, x_3\in(-\infty,+\infty)$, for $t\in\mathbb{R}$ \cite{Ali-Akbari:2017vtb}, as illustrated by Fig. \ref{fig:fig}. \vspace*{-0.6cm}

The minimal area of $\upgamma_A$, which is proportional to the HEE of $A$, is derived when the area in Eq. (\ref{eee}), rewritten as 
\be\label{area} %
 {\scalebox{0.84}{$\textsc{Area}$}}(\upgamma_A)=\frac{1}{4G_5}\int \sqrt{|g_{\upgamma_A}|}\,dx_1dx_2dx_3,
\ee 
is minimized, where $g_{\upgamma_A}$ is the determinant of the induced metric on $\upgamma_A$. Therefore the HEE reads 
\beq\label{EE1}
\!\!\!\!\!\!\!\!S_A&=&\frac{A_2}{4G_5}\bigintsss_{\,-\frac{\ell}{2}}^{\frac{\ell}{2}} \sqrt{\frac{z_0^6}{z^6}+\frac{z_0^6}{z^6C(z)} z^{\prime 2}
}\,dx_1,
\eeq
where $A_2$ denotes the area of the 2-dimensional surface generated by $(x_2, x_3)$, whereas the notation $z'=\frac{dz}{dx_1}$ is also used. Therefore, employing Eq. \eqref{metric1} yields\footnote{Hereon $R=1$ is adopted for simplicity, however all numerical calculations that follows take it into account.}
\beq\label{EE2}
\!\!\!\!\!\!\!\!\!\!\!\!\!S_A&=&\frac{A_2}{4G_5}\!\bigintsss_{-\frac{\ell}{2}}^{\frac{\ell}{2}} \left[\frac{z_0^2z^{\prime 2} \left((1\!-\!4 \upbeta ) \frac{z^4}{z_0^4}\!+\!2\right)}{\frac{z^4}{z_0^4} \left( \frac{3z^8}{z_0^8}\!-\!
 \frac{5z^4}{z_0^4}\!+\!2\right)}\!+\!\frac{z_0^6}{z^6}\right]^{1/2}\!\!\!\!\!\!dx_1.
\eeq
As the area does not explicitly depend on $x_1$, the Hamiltonian is a constant of motion, 
\beq\label{derevative}
\!\!\!\!\!\!\!\!\!\!\frac{{}\sqrt{{R^2 z^2C^2(z)}+{z^{\prime 2}}}}{zC(z)R^4}&=&{\sqrt{\frac{ \frac{z^{\prime 2}}{z_0^2} \left((1\!-\!4 \upbeta ) \frac{z^4}{z_0^4}+2\right)}{ \frac{3z^8}{z_0^8}\!-\! \frac{5z^4}{z_0^4}\!+\!2}\!+\!1}}
\eeq
that equals ${z_\star^3}$, 
for $\displaystyle{\lim_{x_1\to0}z'=0}$. Hence, Eq. \eqref{derevative} yields the following ODE,  
\beq\label{zprime}
z^{\prime2}&=&{\frac{\left(2- \frac{3z^4}{z_0^4}\right) \left(1-\frac{z^4}{z_0^4}\right) \left(\frac{z_\star^6}{z^6}-1\right)}{(1-4
 \upbeta ) \frac{z^4}{z_0^4}+2}}.
\eeq
Therefore the diametral size of the minimal surface $\upgamma_A$, associated with the deformed black brane setup, is given by 
\beq
\ell&=&
2\bigintsss_0^{z_\star}\sqrt{\frac{z_\star^6}{z^6}+\frac{(4 \upbeta -1) \frac{z^4}{z_0^4}-2}{ \frac{3z^8}{z_0^8}-\frac{5 z^4}{z_0^4}+2}}\,{dz}.
\eeq
When Eq. \eqref{zprime} is replaced into \eqref{EE1} one obtains the HEE, 
\beq\label{entanglement}
\!\!\!\!\!\!\!\!\!\!\!\!S_A
&=&\frac{A_2}{2G_5}\bigintsss_{\,0}^{z_\star}\!\!\!\frac{(1-4 \upbeta ) \frac{z^4}{z_0^4}+2}{\frac{z^7}{z_0} \left(\frac{3 z^8}{z_0^8}-\frac{5 z^4}{z_0^4}+2\right)
 \left(\frac{1}{z^6}-\frac{1}{z_\star^6}\right)}\,dz.
\eeq

\begin{figure}[H]
\begin{center}
\includegraphics[width=82 mm]{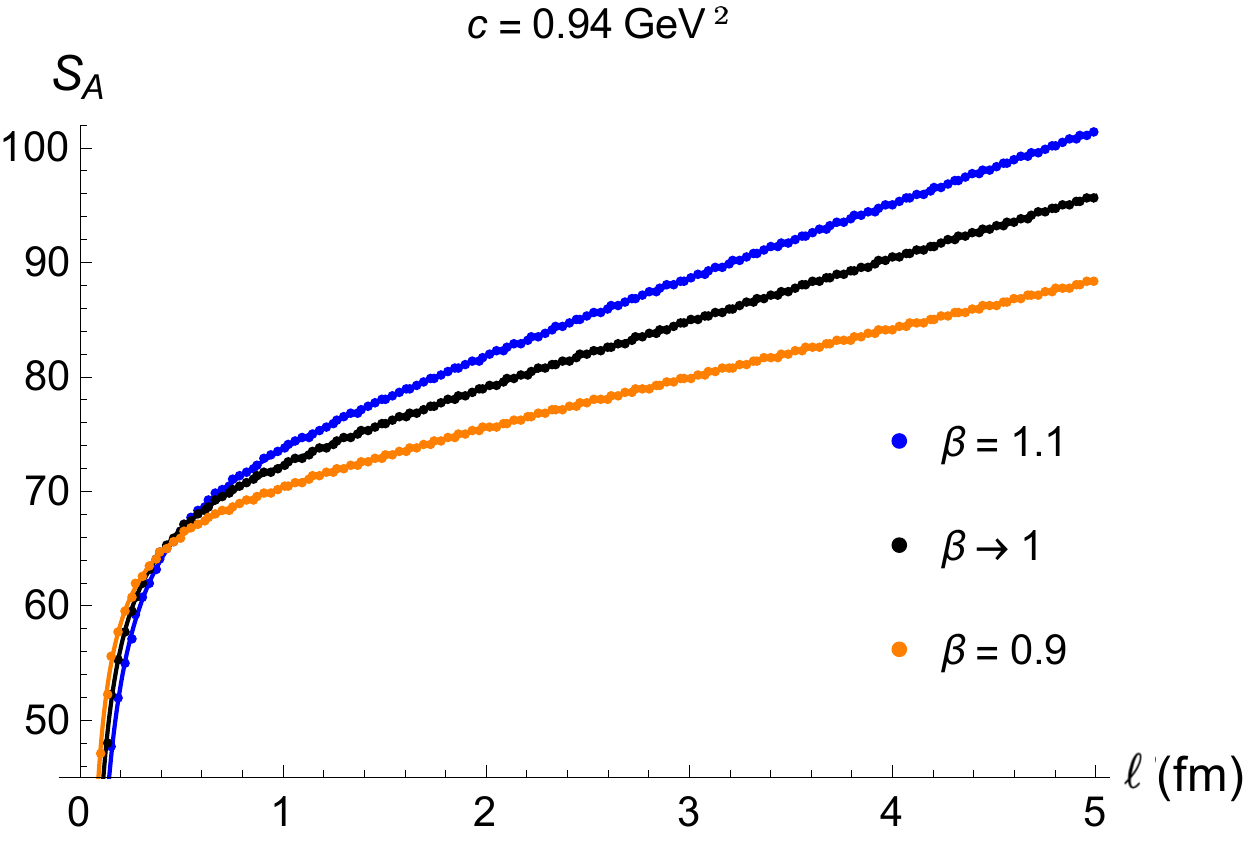}
\caption{\small{HEE $S_A(\ell)$ in terms of $\ell$, for $c= 0.94$ ${\rm GeV}^2$. Numerical results are plotted as orange points, for $\upbeta = 0.9$; as black points, for $\upbeta \to 1$; and as blue points, for $\upbeta = 1.1$, respectively interpolated by the respective lines. }} \label{222}
\end{center}
\end{figure}

\begin{figure}[H]
\begin{center}
\includegraphics[width=82 mm]{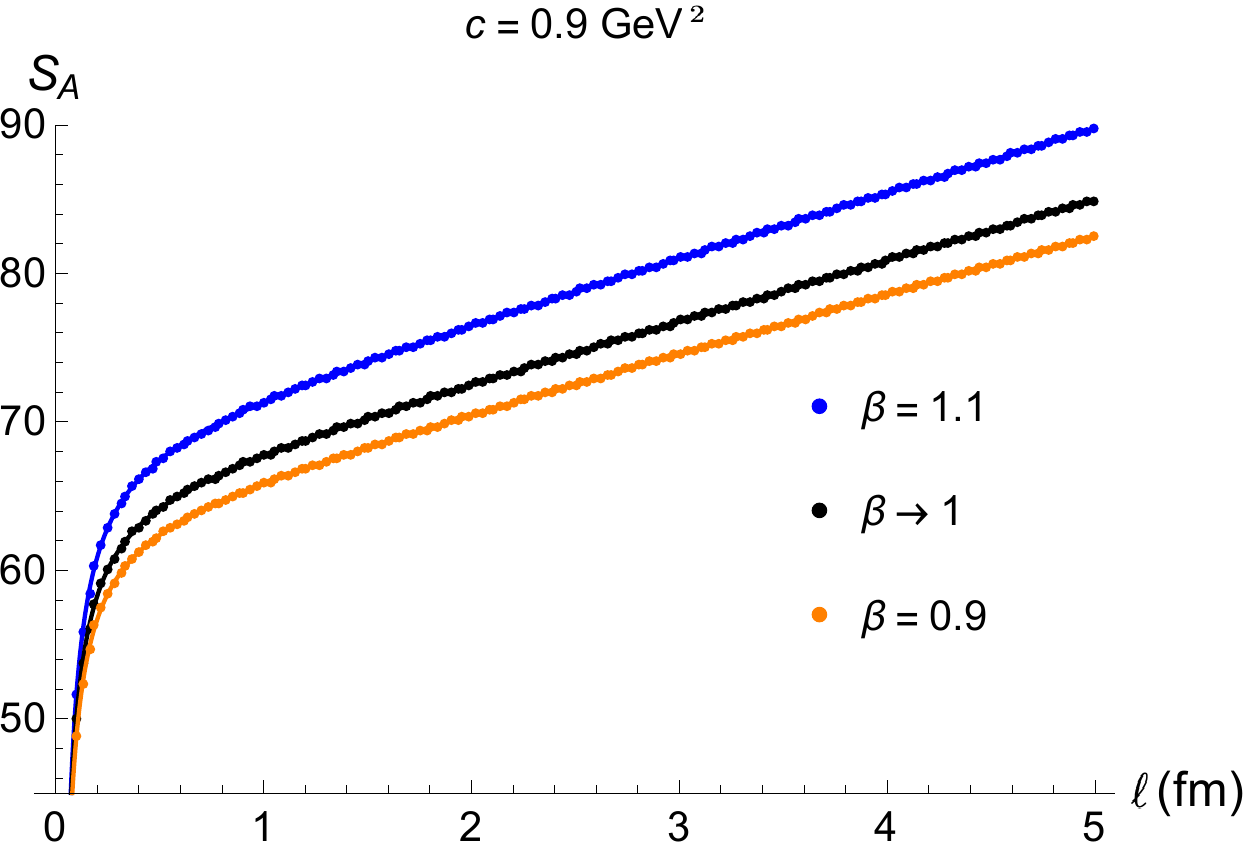}
\caption{\small{HEE $S_A(\ell)$ in terms of $\ell$, for $c= 0.9$ ${\rm GeV}^2$. Numerical results are plotted as orange points, for $\upbeta = 0.9$; as black points, for $\upbeta \to 1$; and as blue points, for $\upbeta = 1.1$, respectively interpolated by the respective lines.} } \label{2221}
\end{center}
\end{figure}

\begin{figure}[H]
\begin{center}
\includegraphics[width=82 mm]{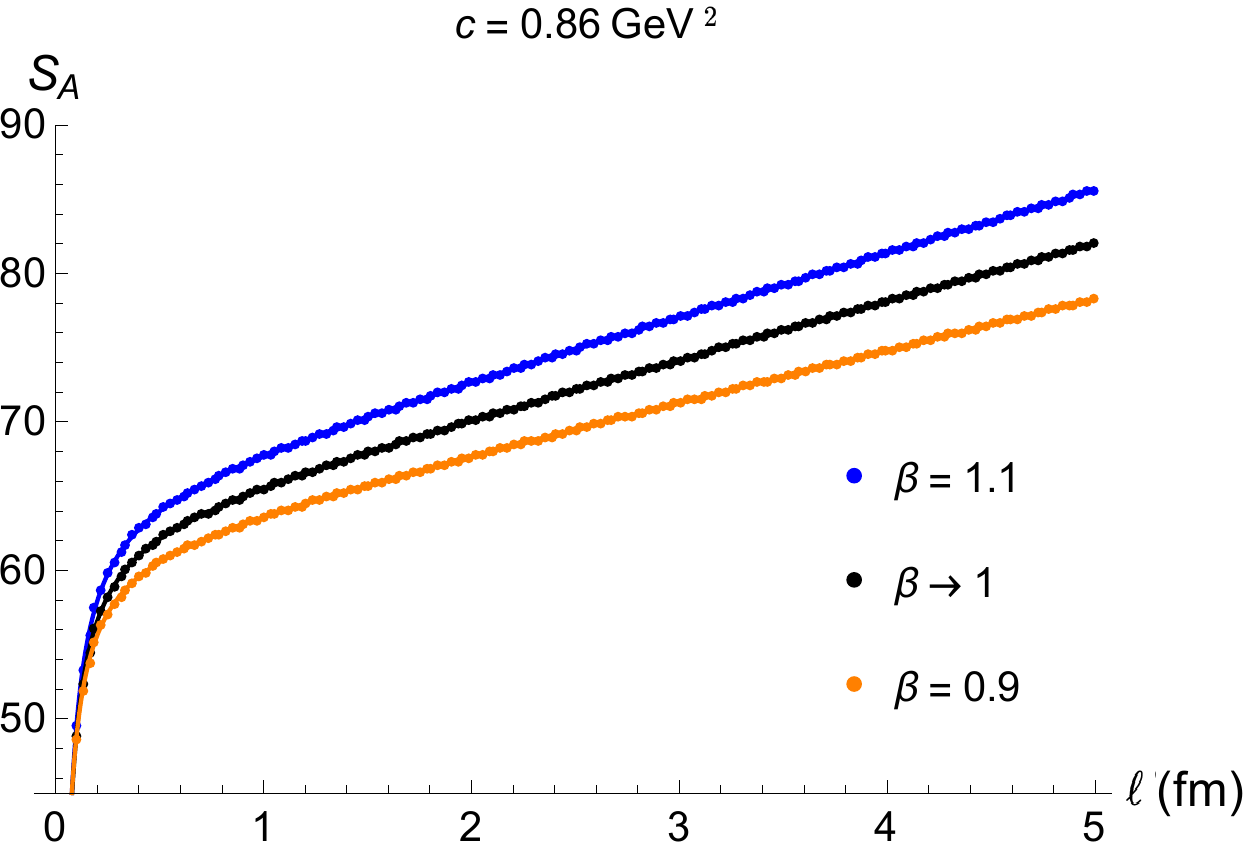}
\caption{\small{HEE $S_A(\ell)$ in terms of $\ell$, for $c\approxeq 0.9$ ${\rm GeV}^2$. Numerical results are plotted as orange points, for $\upbeta = 0.9$; as black points, for $\upbeta \to 1$; and as blue points, for $\upbeta = 1.1$, respectively interpolated by the respective lines.} } \label{2223}
\end{center}
\end{figure}

Figs. \ref{222} -- \ref{2223} display the HEE in terms of $\ell$, for several values of $\upbeta$, with $c$ chosen to represent physically realistic values that match hadronic Regge trajectories.
\noindent For the case $c= 0.94$ ${\rm GeV}^2$, numerical data in Fig. \ref{222} can be interpolated by the functions, respectively for $\upbeta=1.1$, $\upbeta\to1$, and $\upbeta = 0.9$,  
\begin{subequations} 
\beq
S_A(\ell) &=
& -\frac{2.9639}{\ell}+70.0961+5.2789 \,\ell,\;\;\;\label{eeb1}\\ S_A(\ell)&=& -\frac{3.9418}{\ell}+71.4559+6.1637\, \ell,\label{eeb2}\\
S_A(\ell) &=&-\frac{2.1763}{\ell}+68.4564+\,4.0781 \ell.\label{eeb3}
\eeq
\end{subequations}
\noindent Now, for $c= 0.9$ ${\rm GeV}^2$, the plots in Fig. \ref{2221} can be respectively interpolated for $\upbeta=1.1$, $\upbeta\to1$, and $\upbeta = 0.9$,
\begin{subequations} 
\beq
S_A(\ell) &=
& -\frac{1.7639}{\ell}+68.8234+4.2610 \,\ell,\label{1eeb1}\\ S_A(\ell)&=& -\frac{1.5641}{\ell}+65.2788+3.9985\, \ell,\label{1eeb2}\\
S_A(\ell) &=&-\frac{1.5130}{\ell}+63.3537+\,3.8716 \ell.\label{1eeb3}
\eeq
\end{subequations}
The last case to be addressed here is $c=0.86$ ${\rm GeV}^2$.
\noindent Numerical data in Fig. \ref{2223} can be interpolated by the respective functions, for $\upbeta=1.1$, $\upbeta\to1$, and $\upbeta = 0.9$,
\begin{subequations} 
\beq
S_A(\ell) &=
& -\frac{1.6087}{\ell}+65.1094+4.1724 \,\ell,\label{2eeb1}\\ S_A(\ell)&=& -\frac{1.4651}{\ell}+63.9810+3.8453\, \ell,\label{2eeb2}\\
S_A(\ell) &=&-\frac{1.3108}{\ell}+61.4564+\,3.4195 \ell.\label{2eeb3}
\eeq
\end{subequations}

Besides, the non-connected configuration can be set by two disconnected manifolds \beq
\!\!\!\!\!\!\!\!\!\!\!\!M_\pm=\left\{\!\begin{pmatrix}
      x_{1} \\
      x_{2} \\
           x_{3}
     \end{pmatrix}\!\in\mathbb{R}^3\,\Big\vert\,x_1\!=\!\pm \ell/2; \;x_2,x_3\in(-\infty,\infty)\right\}\eeq whose HEE reads 
\beq
\mathring{S}_A&=&\frac{A_2}{2G_5}\int_{0}^{z_\star}\frac{R^3z_0^3}{\sqrt{C(z)}z^3}dz\nonumber\\&=&\frac{A_2}{2G_5}\bigintsss_{\,0}^{z_\star}\frac{z_0^2}{z^2}\sqrt{\frac{(1-4 \upbeta ) \frac{z^4}{z_0^4}+2}{\frac{z^2}{z_0^2} \left(\frac{3z^8}{z_0^8}-\frac{5 z^4}{z_0^4}+2\right)}}\,dz,
\eeq
Introducing the difference of the HEE computed with respect to the connected and disconnected regions, 
\be \label{ds}
\Updelta S(\ell):=\frac{2G_5}{A_2}\left(S_A-\mathring{S}_A\right),
\ee %
the critical length $\ell_c$ is defined as the critical distance such as $\Updelta S(\ell_c)=0$. Then for $\ell\lessgtr\ell_c$, corresponding to $\Updelta S\lessgtr0$, Ref. \cite{Klebanov:2007ws} showed that the HEE varies as a function of the number of colors, $N_c$, in the gauge theory, respectively as $\sim 1$ ($\sim N_c^2$). Thus, there is a deconfinement first order phase transition in the conformal field theory at the boundary, at $\ell=\ell_c$.

\begin{figure}[h]
 \centering
 \begin{subfigure}[b]{0.3\textwidth}
 \centering
 \!\!\!\!\!\!\!\!\!\!\!\!\!\!\! \includegraphics[width=1.5\textwidth]{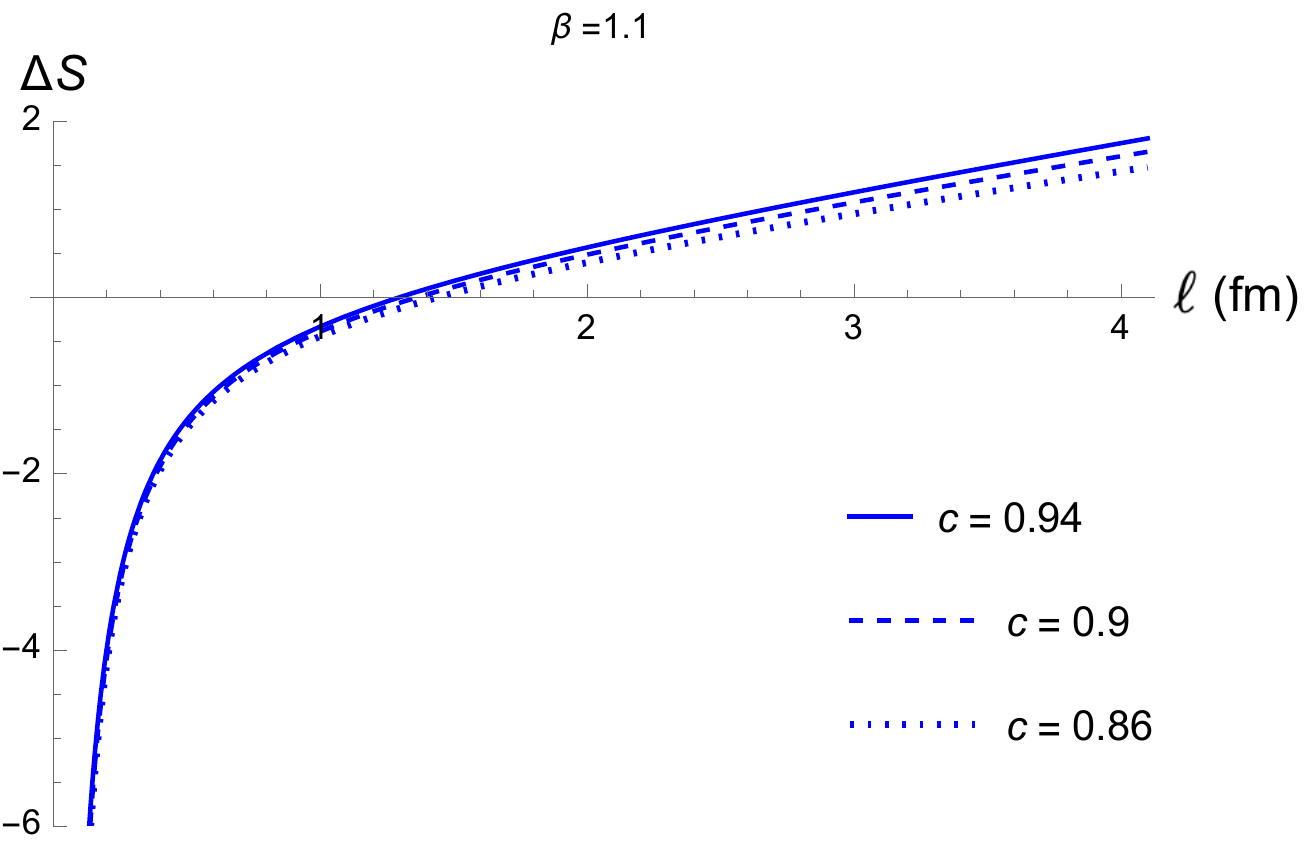}
 \caption{\small{$\upbeta = 1.1$.}}
 \label{fig1b}
 \end{subfigure}\qquad\qquad\qquad\qquad
 \begin{subfigure}[b]{0.3\textwidth}
 \centering
 \includegraphics[width=1.5\textwidth]{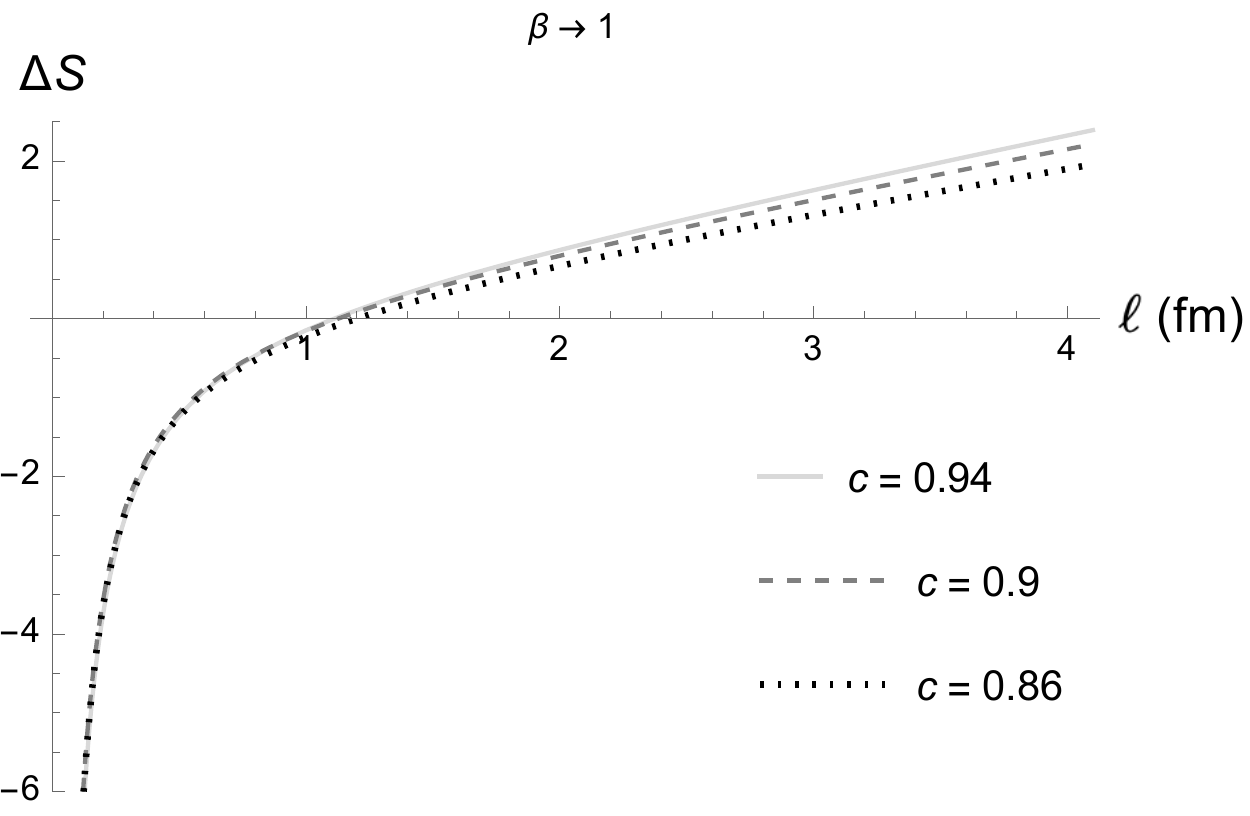}
 \caption{\small{$\upbeta \to1$.}}
 \label{fig1c}
 \end{subfigure}\newline
 \hfill
 \begin{subfigure}[b]{0.3\textwidth}
 \centering
 \includegraphics[width=1.5\textwidth]{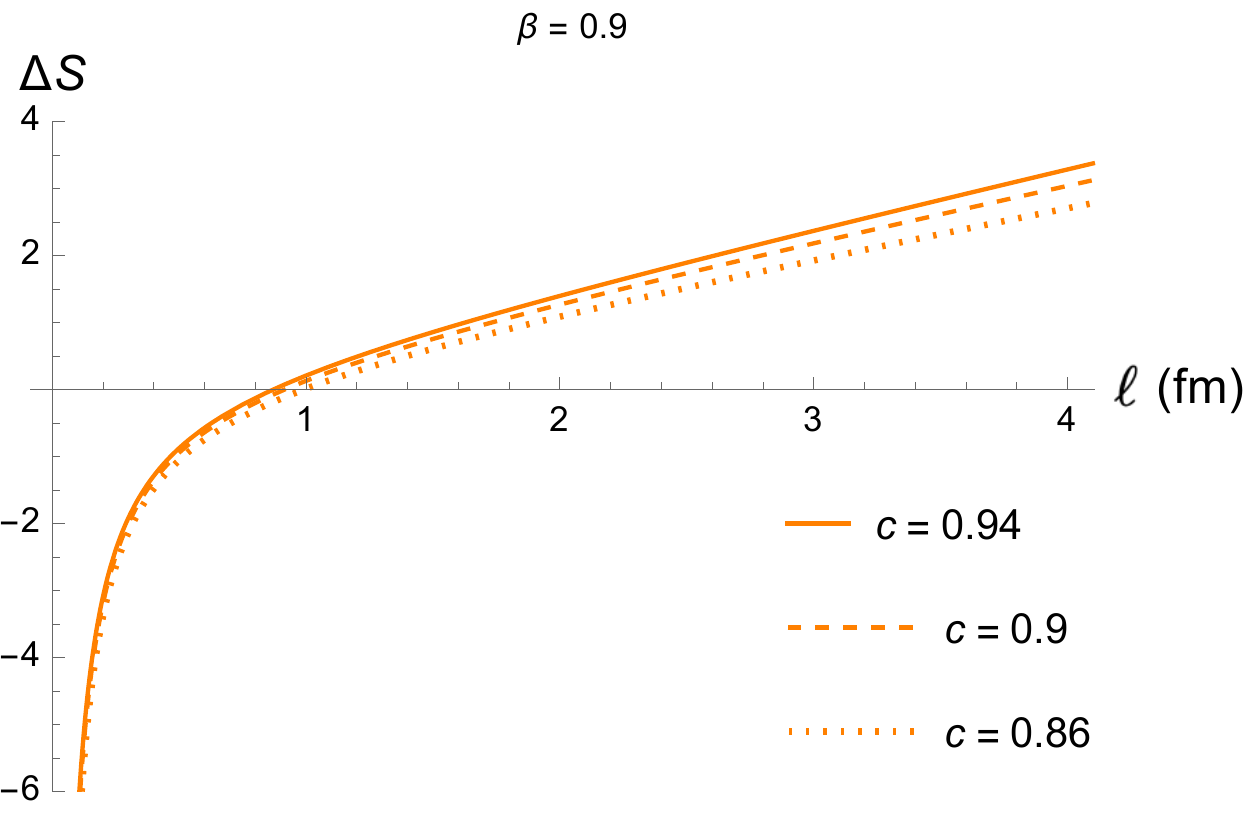}
 \caption{\small{$\upbeta = 0.9$.}}
 \label{fig1d}
 \end{subfigure}
 \caption{\small{$\Delta S(\ell)$ in terms of $\ell$ for three different values of $c$. The continuous lines depict the case where $c=0.94$ GeV${}^2$ and the dashed lines display the case where $c=0.9$ GeV${}^2$, whereas the dotted lines illustrate the case where $c=0.86$ GeV${}^2$, for each fixed value of $\upbeta$.}}
 \label{fig100}
\end{figure}

In the deformed black brane setup given by the metric (\ref{metric1}, \ref{eq:Nu}, \ref{eq:Au}), the potential energy \eqref{potential} determines a stable $q\bar{q}$ confined bound state. Figs. \ref{fig1b} -- \ref{fig1d} illustrate the variation $\Updelta S$ of the HEE between the connected and the disconnected regions, for several values of $c$ and $\upbeta$. The quantity $\Updelta S$ in Eq. (\ref{ds}) determines a phase transition between confined and deconfined phases. In particular, when $\ell>\ell_c$ the system lives in a confined phase \cite{Rougemont:2017tlu}.
\noindent The plots in Fig. \ref{fig1b} can be numerically determined, for $\upbeta =1.1$, respectively for $c=0.86$ GeV${}^2$, $c=0.9$ GeV${}^2$, and for $c=0.94$ GeV${}^2$, by 
\begin{subequations} 
\beq
\Updelta S(\ell) &=& -\frac{0.8512}{\ell}+0.4120 \ell,\label{b1}\\
\Updelta S(\ell) &=& -\frac{0.8388}{\ell}+0.4539 \ell,\label{b2}\\
\Updelta S(\ell) &=& -\frac{0.8182}{\ell}+0.4897 \ell.\label{b3}
\eeq
\end{subequations}
Besides, the plots in Fig. \ref{fig1c} can be numerically interpolated, for $\upbeta\to1$, respectively for $c=0.86$ GeV${}^2$, $c=0.9$ GeV${}^2$, and for $c=0.94$ GeV${}^2$, by the following expressions,
\begin{subequations} 
\beq
\Updelta S(\ell) &=& -\frac{0.7412}{\ell}+0.5837 \ell,\label{c1}\\
\Updelta S(\ell) &=& -\frac{0.7538}{\ell}+0.5220 \ell,\label{c2}\\
\Updelta S(\ell) &=& -\frac{0.7782}{\ell}+0.6294 \ell,\label{c3}
\eeq
\end{subequations}
whereas the graphics in Fig. \ref{fig1d} for $\upbeta=0.9$ have the following expressions, numerically determined, respectively for $c=0.86$ GeV${}^2$, $c=0.9$ GeV${}^2$, and for $c=0.94$ GeV${}^2$, by  
\begin{subequations} 
\beq
\Updelta S(\ell) &=& -\frac{0.6538}{\ell}+0.8620 \ell,\label{d1}\\
\Updelta S(\ell) &=& -\frac{0.6725}{\ell}+0.8024 \ell,\label{d2}\\
\Updelta S(\ell) &=& -\frac{0.7085}{\ell}+0.7206 \ell.
\eeq
\end{subequations}
Therefore, the values of the critical lencth $\ell_c$, defined as $\Updelta S(\ell_c)=0$ are shown in Table \ref{tab1}, for the respective values of $c$ and $\upbeta$.
 \begin{table}[H]
\centering
\begin{tabular}{||c||c|c|c||}
\hline\hline
\,\diagbox{$\upbeta$}{$c$ (GeV${}^2$)} \,&\, 0.86 \,&\, 0.90 \,&\, 0.94\,\\
\hline\hline
$1.1$\,&\, 1.4373 \,&\, 1.3594 \,&\, 1.2926\,\\\hline
$1$\,&\, 1.2017 \,&\, 1.1269 \,&\, 1.1119\,\\\hline
$0.9$\,&\, 0.9914 \,&\, 0.9157 \,&\, 0.8709\,\\\hline
\hline
\end{tabular}
\caption{\small{Critical length $\ell_c$ (fm), for several values of $c$ and $\upbeta$.}}\label{tab1}
\end{table}
\noindent Table \ref{tab1} shows a phase transition that takes place when $\ell_c$ has around 1 fm order of magnitude, for the physically realistic values of $c$ and $\upbeta$. These values are compatible with the separation between a quark and an antiquark in a $q\bar{q}$ system.
Phenomenological values $c\sim 0.9$ GeV$^2$ have been employed in the literature. As the plots in Fig. \ref{fig100} show the function $\Updelta S$ as a function of $\ell$, one can realize that for each fixed value of $\upbeta$, the higher the values of the parameter $c$, the shorter the critical length $\ell_c$ is. 

\begin{figure}[H]
\begin{center}
\includegraphics[width=92 mm]{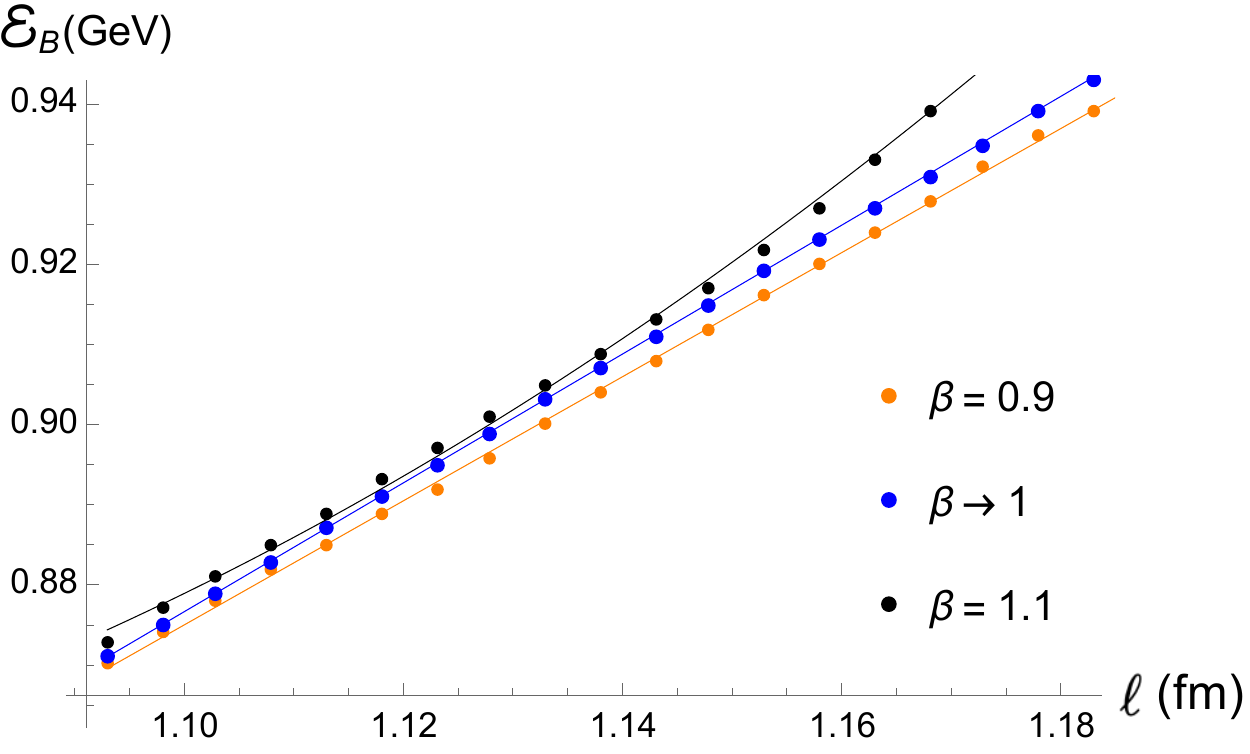}
\caption{\small{Binding energy, $\varepsilon_B$, as a function of the critical length, $\ell_c$. Numerical results are plotted as black points, for $\upbeta = 0.9$; as blue points, for $\upbeta \to 1$; and as orange points, for $\upbeta = 1.1$, respectively interpolated by the respective lines. \label{binding}}}
\end{center}
\end{figure}

The $\bar{q}q$ binding energy can be also derived as $\varepsilon_B\simeq V(\ell_c)$, where the critical length $\ell_c$ drives the bound state maximal size. Fig. \ref{binding} displays the binding energy with respect to $\ell_c$. 
Emulating previous results in the literature \cite{Ali-Akbari:2017vtb}, the binding energy here calculated lies in the range $0.868 GeV < \varepsilon_B\lesssim$ 0.971 GeV, complying to the expected range $0.5\,{\rm GeV} <\varepsilon_B<1$ GeV obtained in Ref. \cite{Satz:2011wf}. 
Numerical data in Fig. \ref{binding} can be interpolated by the respective polynomial functions
\begin{subequations} 
\beq
\varepsilon_B(\ell_c) &=& 0.8043\,\ell_c-8.1521\times 10^{-3},\label{eb1}\\
\varepsilon_B(\ell_c) &=& 0.7747\,\ell_c+2.2770\times 10^{-2},\label{eb2}\\
\varepsilon_B(\ell_c) &=&3.1579\,\ell_c^2 -6.2782\,\ell_c+3.9639,\label{eb3}
\eeq
\end{subequations}
within $0.1\%$, $0.05\%$, and $0.2\%$ root-mean-square deviation, correspondingly. 

Ref. \cite{Andreev:2006eh} proved that deconfinement phase transition takes place when $z_0=z_\star=\sqrt{2/c}$, for the AdS${}_5$--Schwarzschild standard black brane \cite{Casadio:2016aum}. In the case of the deformed black brane (\ref{metric1}, \ref{eq:Nu}, \ref{eq:Au}), $T_c$ can be identified to the deconfinement phase transition critical temperature separating the deconfined quark-gluon plasma to the confined hadronic phase, reading 
\begin{eqnarray} \label{ttt}
	T_c=\frac{1}{\pi}\sqrt{\frac{c(\upbeta-2)}{2(3-4\upbeta)}}.
	\end{eqnarray}
	From the string theory point of view, the flux tubes binding $q$ to $\bar{q}$ break out when the potential energy exceeds the threshold of spontaneous pair formation.
 The values of $T_c$ are shown in Table \ref{tab10}, for $\upbeta =0.9$, $\upbeta \to1$ and $\upbeta =1.1$, for several values of $c$ that are phenomenologically compatible.
 It is a usual procedure to analyze AdS/QCD
predictions and compare them to the deconfinement temperature range $122\,{\rm MeV}\lesssim T_c\lesssim 170$ MeV, as predicted in the AdS/QCD hard-wall model and lattice QCD as well. On the other hand, the AdS/QCD soft-wall model yields the range $190\,{\rm MeV}\lesssim T_c\lesssim 200$ MeV. The value $T_c \approx 175\pm 15$ MeV was derived from analyzing the glueball spectrum \cite{Afonin:2018era}, whereas some other important aspects of the deconfinement
temperature were scrutinized in Ref. \cite{Braga:2020opg,Braga:2019jqg}. The two leading lattice collaborations in this field reported the
values $T_c = 156.5 \pm 9$ MeV and $T_c = 154 \pm 9$ MeV \cite{Borsanyi:2010bp}. 
Experimental data obtained from the Relativistic Heavy Ion Collider (RHIC), the A Large Ion Collider Experiment (ALICE), and the Super Proton Synchrotron (SPS) 
have been discovering features of the quark-gluon plasma, that also describes the early universe \cite{Fernandes-Silva:2018abr}. In fact, the HotQCD Collaboration has found $T_c = 156.5 \pm 1.5$ MeV \cite{HotQCD:2014kol}.

 \begin{widetext}
 \begin{table}[H]
\vspace{1 mm}
\centering
\begin{tabular}{||c||c|c|c|c|c|c|c||}
\hline\hline
\,\diagbox{$\upbeta$}{$c$ (GeV${}^2$)}
 \,&\, 0.86 \,&\, 0.88 \,&\, 0.90 \,&\, 0.92 \,&\, 0.94 \,&\, 0.96 \,&\, 0.98\,\\
\hline\hline
\,$1.1$\,&\, 134.13 \,&\, 135.72 \,&\, 137.24\,&\, 138.81 \,&\, 140.27 \,&\, 141.76\,&\, 143.22\,\\\hline
\,$1$ \,&\, 167.30 \,&\, 169.27 \,&\, 171.18 \,&\, 173.13 \,&\, 174.95 \,&\, 176.80 \,&\, 178.63\,\\\hline
\,$0.9$\,&\, 226.52 \,&\, 229.19 \,&\, 231.78 \,&\, 234.42 \,&\, 236.88 \,&\, 239.88 \,&\, 241.87\,\\\hline
\hline
\end{tabular}
\caption{\small{Critical temperatures (MeV), for several values of $c$ and $\upbeta$.}}\label{tab10}
\end{table}
\end{widetext}
\noindent Table \ref{tab10} illustrates that the parameter $c$ can be more precisely estimated. The HEE is here shown to be a relevant instrument to probe the deconfinement phase transition. The value $c=0.947$ GeV$^2$, when $\upbeta\to1$, represents the most reliable value when one takes into account predictions of the glueball spectrum phenomenology, whereas the value $c=0.982$ GeV$^2$ is compatible to $\upbeta\approxeq0.981$, in the same context. 
Table \ref{tab2} displays the values of $c$ and their corresponding values of $\upbeta$, matching experimental data from HotQCD Collaboration \cite{HotQCD:2014kol}, 
\begin{widetext}
 \begin{table}[H]
\vspace{1 mm}
\centering
\begin{tabular}{||c||c|c|c|c|c|c|c||}
\hline\hline
 \,$c$ (GeV${}^2$)\,&\, 0.86 \,&\, 0.88 \,&\, 0.90 \,&\, 0.92 \,&\, 0.94 \,&\, 0.96 \,&\, 0.98\,\\
\hline
\,$\upbeta$\,&\,$1.031$\,&\, 1.035 \,&\, 1.038 \,&\, 1.043\,&\, 1.048 \,&\, 1.052\,&\,1.057\,\\\hline
\hline
\end{tabular}
\caption{\small{Values of $c$ and corresponding values of $\upbeta$ that match experimental data for the critical temperature, $T_c = 156.5 \pm 1.5$ MeV, from the HotQCD Collaboration \cite{HotQCD:2014kol}.}}\label{tab2}
\end{table}
\end{widetext}
\noindent Besides, the values of $c$ and their corresponding values of $\upbeta$, matching the predictions by the AdS/QCD hard wall model are shown in Table \ref{tab3}. 
\begin{widetext}
{{ \begin{table}[H]
\vspace{1 mm}
\centering
\begin{tabular}{||c||c|c|c|c|c|c||}
\hline\hline
 \,$c$ (GeV${}^2$)\,&\, 0.86 \,&\, 0.88 \,&\, 0.90 \,&\, 0.92 \,&\, 0.94 \,&\, 0.96 \,\\
\hline
\,$[\upbeta_{\scalebox{0.8}{$\textsc{min}$}}, \upbeta_{\scalebox{0.8}{$\textsc{max}$}}]$\,&\,$(1, 1.149]$\,&\, $(1, 1.156]$ \,&\, $[1.001, 1.171]$ \,&\, $[1.007, 1.177]$\,&\, $[1.011, 1.184]$ \,&\, $[1.015, 1.190]$\,\\\hline
\hline
\end{tabular}
\caption{\small{Values of $c$ and corresponding range $[\upbeta_{\scalebox{0.8}{$\textsc{min}$}}, \upbeta_{\scalebox{0.8}{$\textsc{max}$}}]$ that match the AdS/QCD hard wall model.}}\label{tab3}
\end{table}}}
\end{widetext}
\noindent For the soft wall model, the values of $c$ and their corresponding values of $\upbeta$ are shown in Table \ref{tab4}. However, in the context of Eq. (\ref{ra1}), coming from the fact that the shear viscosity-to-entropy density ratio is negative when $0.9\leq \upbeta <1$, there are no physically allowed values of $\upbeta$ that match realistic values of $c$ in QCD such that the critical temperature lies in the range $190\,{\rm MeV}\lesssim T_c\lesssim 200$ MeV, predicted by the soft wall model.
Therefore, the case relating to the soft wall model is here illustrated in Table \ref{tab4} as a matter of completeness, as it has no practical realization for being far from the strictest experimental value $T_c = 156.5 \pm 1.5$ MeV \cite{HotQCD:2014kol}. 
\begin{widetext}
{\small{ \begin{table}[H]
\vspace{1 mm}
\centering
\begin{tabular}{||c||c|c|c|c|c|c||}
\hline\hline
 \,$c$ (GeV${}^2$)\,&\, 0.86 \,&\, 0.88 \,&\, 0.90 \,&\, 0.92 \,&\, 0.94 \,&\, 0.96 \,\\
\hline
\,$[\upbeta_{\scalebox{0.8}{$\textsc{min}$}}, \upbeta_{\scalebox{0.8}{$\textsc{max}$}}]$\,&\,$[0.936, 0.953]$\,&\, $[0.941, 0.956]$ \,&\, $[0.946, 0.961]$ \,&\, $[0.95, 0.967]$\,&\, $[0.954, 0.972]$ \,&\, $[0.958, 0.976]$\,\\\hline
\hline
\end{tabular}
\caption{\small{Values of $c$ and corresponding range $[\upbeta_{\scalebox{0.8}{$\textsc{min}$}}, \upbeta_{\scalebox{0.8}{$\textsc{max}$}}]$ that match the AdS/QCD soft wall model.}}\label{tab4}
\end{table}}}
\end{widetext}

%Table \ref{tab3} [\ref{tab4}]. 
%{\small{ \begin{table}[H]
%\vspace{1 mm}
%\centering
%\begin{tabular}{||c||c|c|c|c|c|c||}
%\hline\hline
 %\,$c$ (GeV${}^2$)\,&\, 0.86 \,&\, 0.88 \,&\, 0.90 \,&\, 0.92 \,&\, 0.94 \,&\, 0.96 \,\\
%\hline
%\,$[\upbeta_{\scalebox{0.8}{$\textsc{min}$}}, 5\upbeta_{\scalebox{0.8}{$\textsc{max}$}}]$\,&\,$[0.993, 1.149]$\,&\, $[0.998, 1.156]$ \,&\, $[1.001, 1.171]$ \,&\, $[1.007, 1.177]$\,&\, $[1.011, 1.184]$ \,&\, $[1.015, 1.190]$\,\\\hline
%\hline
%\end{tabular}
%\caption{\small{Values of $c$ and corresponding range $[\upbeta_{\scalebox{0.8}{$\textsc{min}$}}, \upbeta_{\scalebox{0.8}{$\textsc{max}$}}]$ that match the AdS/QCD hard wall model.}}\label{tab3}
%\end{table}}}

\section{conclusions}
\label{c34}

Deformed black branes were used in a gauge/gravity-like duality to analyze the confinement/deconfinement phase transition in QCD at the boundary. Reciprocally, current experimental data regarding QCD can restrict the range of the parameter that rules the deformed black brane, imposing a stricter bound on it. 
The HEE was shown to play the role of the order parameter that controls the confinement/deconfinement phase transition. 
Besides, analyzing the variation of the HEE between connected and disconnected regions yields a critical length at which a deconfinement first-order phase transition in the boundary QCD sets in. 
The HEE for several values of $\upbeta$, for $c=0.86$ GeV${}^2$, $c=0.9$ GeV${}^2$, and $c=0.94$ GeV${}^2$, 
was displayed in Figs. \ref{222} -- \ref{2223}, with important conclusions. Respective interpolation functions were displayed in Eqs. \eqref{eeb1} -- \eqref{2eeb3}, for each fixed value of $\upbeta$, the HEE was shown to increase as $c$ increases. Also, for each fixed value of $c$, the higher the value of $\upbeta$, the more the HEE increases.

The critical length is shown to slightly vary as a function of the deformed black brane parameter, for several values of another parameter (0.86 GeV${}^2$ $\lesssim c\lesssim$ 0.94 GeV${}^2$) in AdS/QCD. 
For lengths above the critical length, the QCD system was shown to reside in a confined phase. 
The results obtained, displayed in Table \ref{tab1}, show a phase transition that takes place when $\ell_c$ has around 1 fm order of magnitude, for the physically realistic values of $c$ and $\upbeta$. These values are compatible with the separation between a quark and an antiquark in a $q\bar{q}$ system. Besides, the plots in Fig. \ref{fig100} show the function $\Updelta S$ with respect to $\ell$. For each fixed value of $\upbeta$, the higher the values of the parameter $c$, the shorter the critical length $\ell_c$ is. Interpolation functions are also presented. 
The binding energy of the quark-antiquark bound state was also addressed, with important results. Fig. \ref{binding} show that for $\upbeta\to 1$ and $\beta = 0.9$, the binding energy is a linear function of the length $\ell$, whereas the case $\beta = 1.1$ can be approximated by a quadratic function of $\ell$, for energies in the range $[0.868$ GeV, 0.971 GeV], which complies to the expected theoretical result. However, contrary to the standard AdS${}_5$--Schwarzschild black brane, where the binding energy is a linear function of the length $\ell$, for the same range of $c$, here in the case of the deformed black brane there is a departure of the linear regime when $\upbeta = 1.1$. The critical temperature at which the deconfinement phase transition that separates the deconfined quark-gluon plasma to the confined hadronic phase was determined and shown to vary with respect to the deformed black brane parameter and $c$. Comparative analysis of theoretical results from the AdS/QCD hard and soft wall models, lattice QCD, the glueball spectrum, and experimental data at RHIC, ALICE, and SPS has been implemented. Finally, deformed black branes in AdS${}_4$ have been introduced in Ref. \cite{Ferreira-Martins:2019wym}, in the context of AdS/CMT. The dual CFT that describes Dirac fluids emulating graphene can be also explored in the context of the HEE. Other generalized black branes can be also addressed \cite{Casadio:2016zhu,Ferreira-Martins:2021cga} and studied in the context of AdS/QCD, including entanglement in other quantum information measures. 

\section*{Acknowledgement}
RdR thanks grants No. 2017/18897-8 and No. 2021/01089-1, São Paulo Research Foundation (FAPESP); and the National Council for Scientific and Technological Development -- CNPq, grants No. 303390/2019-0 and No. 406134/2018-9, and No. 402535/2021-9, for partial financial support.
 
\bibliography{1bib_HEE}

\end{document}